\documentclass[12pt]{iopart}

\def \be{\begin{equation}}
\def \ee{\end{equation}}
\def \bea{\begin{eqnarray}}
\def \eea{\end{eqnarray}}

\def \s2{\sqrt 2}

\def \vol{\sqrt{-g}\,}

\begin{document}

\jl{6}

\title{Analysis of relativistic hydrodynamics in conservation form}

\author{Philippos Papadopoulos\dag\footnote[3]{For correspondence:
philippos.papadopoulos@port.ac.uk} and Jos\'e A. Font\ddag}

\address{\dag\ School of Computer Science and Mathematics,
University of Portsmouth, Portsmouth, PO1 2EG United Kingdom}

\address{\ddag\ Max-Planck-Institut f\"ur Astrophysik,
Karl-Schwarzschild-Str. 1, 85740 Garching, Germany}

\begin{abstract}
Formulations of Eulerian general relativistic ideal hydrodynamics in
conservation form are analyzed in some detail, with particular
emphasis to geometric source terms. Simple linear transformations of
the equations are introduced and the associated equivalence class is
exploited for the optimization of such sources. A significant
reduction of their complexity is readily possible in generic
spacetimes. The local characteristic structure of the standard member
of the equivalence class is analyzed for a general equation of state
(EOS). This extends previous results restricted to the polytropic
case. The properties of all other members of the class, in particular
specialized forms employing Killing symmetries, are derivable from the
standard form. Special classes of EOS are identified for both
spacelike and null foliations, which lead to explicit inversion of the
state vector and computational savings. The entire approach is equally
applicable to spacelike or lightlike foliations and presents a
complete proposal for numerical relativistic hydrodynamics on
stationary or dynamic geometries.
\end{abstract}


\submitted

\maketitle

\section{Introduction}
\label{sec:intro}

Relativistic hydrodynamics (RHD) is a basic building block in current
efforts in numerical relativistic astrophysics and relativity. Those
programmes provide much needed theoretical support to observational
efforts focusing on extreme astrophysical systems.  The development of
numerical RHD has followed mostly on the steps of numerical {\em
non}-relativistic hydrodynamics, whose technological significance has
prompted the generation over the past decades of a large amount of
mathematical and algorithmic known-how. This knowledge has been
historically transferred effectively to the relativistic case, the
prime example being artificial viscosity techniques~\cite{av,rm67},
which have been adopted and advocated by the pioneering work of
Wilson~\cite{wilson72}.

More recent techniques are based on deeper mathematical understanding
of non-linear conservation laws~\cite{godunov,lax,leveque}. The
relativistic Euler equations are also readily analyzed in conservation
form (see~\cite{anile} for a modern review of related mathematical
aspects). Numerical formulations of conservative RHD were first
presented in~\cite{mim} for the one-dimensional case. A
multi-dimensional extension, using an explicit ``3+1'' decomposition
of spacetime, and valid for a general equation of state (EOS), was
given in~\cite{betal97}. In~\cite{em95} a covariant formulation was
presented, adapted to polytropic EOS and a specific numerical solution
procedure (Roe solver). An alternative formulation, restricted to the
special relativistic case was given in~\cite{Falle}. A wide collection
of numerical applications based on those approaches is reviewed
in~\cite{jcam}.

Adopting the point of view of non-linear conservation laws liberates
the analysis from the need to adhere to Newtonian fluid dynamical
concepts. This seems to be a natural approach when using RHD in
general relativistic studies, as is exemplified by the ease with which
one can extend RHD methods e.g.,to lightlike foliations of spacetimes
- a highly non-Newtonian concept. In~\cite{pf99} we introduced a new
covariant approach, significantly simpler than~\cite{em95}, but
similarly restricted to perfect fluid EOS. This formulation was
extensively tested numerically, in spherical symmetry, in particular in
the context of dynamical black hole accretion~\cite{pf99}. The main new 
results in the present paper extend this formulation to a general EOS and
introduce a simple linear analysis of the RHD equations which allows
the tailoring of geometric source terms to situations of interest.
The discussion unifies and clarifies apparently disconnected choices
the characterize the literature mentioned above and will hopefully
assist the future development of the field.

The paper is organized as follows: In 
Section~\ref{sec:formalism} we introduce the useful notion of linearly
equivalent conservation laws and in Section~\ref{sec:rhd} we show how
the use of the associated transformations allows to tailor the
geometrical source structure. In particular we develop there a form of
the equations suitable for spacetimes with exact or approximate
Killing symmetries. In Section~\ref{sec:analysis} the characteristic
structure of the standard form of the RHD equations is derived for a
general EOS. The inversion of the state vector is discussed in some
detail. Some classes of EOS which lead to explicit inversions are
outlined for the algebraically distinct cases of spacelike and null
foliations. Numerical applications will be presented elsewhere. Some
discussion of related issues is included in Section~\ref{sec:summary}.

Geometrized units ($G=c=1$) are used throughout. The metric sign
conventions follow~\cite{MTW}. Spacetime indices are denoted by small
Greek letters and run from zero to three. Small Latin indices denote
hypersurface coordinates and run from one to three.  Boldface letters
(e.g., ${\bf F}$) denote vectors in the fluid state space, which has
dimension five.  Those dimensions are labeled from zero to four,
coinciding with the spacetime dimension for values up to
three. Partial derivatives with respect to a coordinate $x$ are
denoted as $F_{,x}$ and the summation convention is used.

\section{Equivalence classes of conservation laws}
\label{sec:formalism}

The general form of an N-dimensional system of conservation laws,
expressed in a coordinate system $x^{\mu}=(x^{0},x^{j})$, where
$x^{j}$ parametrize the hypersurfaces of constant time $x^{0}$, is
\begin{equation}
 {\bf U}\!_{,x^{0}} +  {\bf F}^{j} \!_{,x^{j}} = {\bf S} \, , 
\label{eq:cons-law}
\end{equation}
where ${\bf U}$ is an N-vector describing the state of the system, the
flux vectors ${\bf F}^{j}(x^{\mu},{\bf U})$ control the time rate of
change of the system state within an elementary volume, and possible
(conservation violating) source terms are denoted by ${\bf
S}(x^{\mu},{\bf U})$. The independence of the source vector on the state
vector derivatives is worth stressing.

We introduce a $N\times N$ dimensional square matrix ${\bf
G}(x^{\mu})$, which is invertible ($det({\bf G})\not=0$). We consider
the new state variables ${\bf \bar{U}} = {\bf G}{\bf U}$, satisfying
the equations
\begin{equation}
{\bf \bar{U}}\!_{,x^{0}} + {\bf \bar{F}}^{j}\!_{,x^{j}} = {\bf
\bar{S}}^{} + {\bf {G}}\,_{,x^{0}} {\bf {G}}^{-1} {\bf \bar{U}} + {\bf
{G}}\,_{,x^{j}} {\bf {G}}^{-1} {\bf \bar{F}}^{j} \, ,
\label{eq:cons-law-trans}
\end{equation}
where ${\bf \bar{F}} = {\bf G}{\bf F}$ and ${\bf \bar{S}} = {\bf
G}{\bf S}$. The linear transformation introduced by ${\bf G}$ leaves
the characteristic structure of the original system intact.  This can
be seen immediately by writing the homogeneous version of
system~(\ref{eq:cons-law}) in quasi-linear form,
\begin{equation}
{\bf U}\!_{,x^{0}} + {\bf B}^{j}
{\bf U}\!_{,x^{j}} = 0 \, ,
\label{eq:quasi-linear}
\end{equation}
where
\begin{equation}
{\bf B}^{j} =  \frac{ \partial {\bf F}^{j} } { \partial {\bf U}}
\, , 
\label{eq:jacobian}
\end{equation}
are the Jacobians of the flux vectors with respect to the state
vector. The eigenvalues of each Jacobian are then determined by the equation
\begin{equation}
 det({\bf B}^{j} - \lambda^{j} {\bf I}) =  0 \, ,
\end{equation}
where $\lambda^{j}$ denotes an eigenvalue in the direction $j$ and
${\bf I}$ denotes the unit matrix. The corresponding right and left
eigenvectors, ${\bf r}^{j}$ and ${\bf l}^{j}$, are determined by the
equations
\begin{eqnarray}
( {\bf B}^{j} - \lambda^{j} {\bf I}) {\bf r}^{j} & = & 0 , \\
  {\bf l}^{j} ( {\bf B}^{j} - \lambda^{j} {\bf I}) & =& 0 \, .
\end{eqnarray}

It follows then from elementary properties of matrices and
determinants that the transformed system~(\ref{eq:cons-law-trans}) has
along the $j$-th direction the same set of eigenvalues $\lambda^{j}$ as
the original system. The new right eigenvectors given by ${\bf
\bar{r}} = {\bf G}{\bf r}$, while the new left eigenvectors are given
by ${\bf \bar{l}} = {\bf G}^{-1}{\bf l}$.

Smooth solutions ${\bf U}(x^{\mu})$ of the system of
equations~(\ref{eq:cons-law}) obviously lead to smooth solutions
$\bar{{\bf U}}(x^{\mu})$ of the transformed system.  The same is true
for weak solutions, as can easily be seen (in the one-dimensional case)
from the definition of such solutions~\cite{leveque} as those
satisfying the integral relation:
\begin{equation}
\int^{+\infty}_{0} dx^{0} \int^{+\infty}_{-\infty} dx^{1}
\left[ \phi_{,x^0} {\bf U}
+ \phi_{,x^1} {\bf F}^{1} \right] = \int^{+\infty}_{-\infty}
dx^{1} \phi(0,x^1) {\bf U} \, ,
\end{equation}
where $\phi(x^{\mu})$ are continuously differentiable test functions
of compact support. Pre-multiplying this system of integral
conservation laws by ${\bf G}$ immediately shows that weak solutions
are also transformed properly.  Hence from the point of view of a
hyperbolic set of partial differential equations, all such linearly
related systems are equivalent.

Proceeding to the effects of the linear transformation on the source
terms, we note that the point-wise properties of ${\bf G}$ simply
reshuffle the source components among the equations, but the spacetime
dependence of ${\bf G}$ modifies the RHS of the equations in a
non-trivial way. Genuine source terms, e.g., such as those introduced
in Newtonian hydrodynamics to model combustion processes, depend only
on the state vector and cannot be essentially modified in this
manner. In contrast, the regular source terms of RHD are highly
coordinate dependent and involve a variety of metric derivatives.  It
becomes immediately apparent that an appropriate choice can maximize
the conservation nature of the system.

\section{Relativistic fluid dynamics}
\label{sec:rhd}

\subsection{Reductions of the equations}

The above discussion is applicable to a variety of relativistic
systems but we specialize here to RHD. We hence assume a matter
current and stress energy tensor corresponding to a perfect fluid,
i.e., $J^{\mu} = \rho u^{\mu} $, $ T^{\mu\nu} = \rho h u^{\mu} u^{\nu}
+ p g^{\mu\nu}$. In these definitions, $u^{\mu}$ is the fluid four
velocity, which is constrained by the normalization condition
$u^{\mu}u_{\mu}=-1$, $\rho$ is the rest mass density, $p$ is the
pressure, $\varepsilon$ is the specific internal energy and $h = 1 +
\varepsilon + p/\rho$ is the specific enthalpy. The pressure is
determined by a general two-parameter family EOS,
$p=p(\rho,\varepsilon)$. The relativistic conservation equations in
covariant form are given by
\begin{eqnarray}
\label{eq:stress-energy}
\nabla_{\mu} T^{\mu\nu} & = &0 \, , \\
\nabla_{\mu} J^{\mu} & = & 0 \, .
\label{eq:continuity}
\end{eqnarray}

\subsubsection{The standard form of RHD}

Upon introducing a coordinate system, $(x^{0},x^{i})$, one obtains a
simple form for the continuity equation
\begin{equation}
(\sqrt{-g} J^{\mu})\,_{,\mu}  =  0  \, .
\label{eq:continuity2}
\end{equation}

The standard form of the divergence of the stress-energy tensor
can be written as
\begin{equation}
(\sqrt{-g}\, T^{\mu\nu})\,_{,\mu} =  -
 \Gamma^{\nu}_{\mu\lambda} (\sqrt{-g} \, T^{\mu\lambda}) \, ,
\label{eq:upper}
\end{equation}
where
\begin{equation}
\Gamma^{\nu}_{\mu\lambda} = \frac{1}{2} g^{\nu\rho}
(g_{\mu\rho,\lambda} + g_{\lambda\rho,\mu} - g_{\mu\lambda,\rho}) \, ,
\end{equation}
are the usual Christoffel symbols of the second kind. The set of
equations~(\ref{eq:continuity2},\ref{eq:upper}) form a system of
non-linear coupled conservation laws which has been the basis for the
formulations proposed in~\cite{em95,Falle}, and a ``3+1'' decomposition of
the same equations forms the basis of~\cite{betal97}. We hence call
this set of equations the {\em standard form} and will analyze it
further in the sequel. Linear equivalence with many other forms makes
our analysis applicable to a wider set.

The state vector is given by
\begin{eqnarray}
\label{eq:conserved_A1}
{\bf U}^{\mu} & = & \vol T^{0\mu} = \vol (\rho h u^{0} u^{\mu} + p g^{0\mu}) \, , \\
{\bf U}^{4}   & = & \vol J^{0}  = \vol \rho u^{0} \, ,
\label{eq:conserved_A}
\end{eqnarray}
while the flux vectors are given by
\begin{eqnarray}
{\bf F}^{j\mu} & = & \vol T^{j\mu} = \vol (\rho h u^{\mu} u^{j} + p g^{\mu j} )\, , \\
{\bf F}^{j4}   & = & \vol J^{j}    = \vol \rho u^{j} \, ,
\label{eq:fluxes_A}
\end{eqnarray}
and the geometric source terms are (in this case ${\bf F}^{j0}={\bf U}^{0}$)
\begin{eqnarray}
\label{eq:sources_A}
{\bf S}^{\mu} & = & - \, (
    \Gamma^{\mu}_{00} {\bf U}^{0} + 2 \Gamma^{\mu}_{k0} {\bf U}^{k} 
+   \Gamma^{\mu}_{kl} {\bf F}^{kl} ) \, , \\
{\bf S}^{4} & = & 0 \, .
\end{eqnarray}

We note that in RHD the source terms depend on the coordinates, the
state vector {\em and} the flux vector. The dependence of the latter
on the state vector is implicit, as will be elaborated on later.

\subsubsection{The Killing form of the RHD equations}

It is well known that whereas the continuity
equation~(\ref{eq:continuity}) is a true conservation law, the
divergence of the stress-energy tensor will lead in general to true
conservation only when a spacetime symmetry is present~\cite{wald}.
Spacetimes with exact symmetries (e.g., the Kerr spacetime) or
approximate ones (e.g., quasistationary binary configurations with
helicoidal symmetry~\cite{meudon-binaries}) are of wide interest in
fluid-dynamical studies. It is hence of some importance to identify ways
to maximize conservation in this context.

A spacetime symmetry is captured by a Killing vector $\xi_{\nu}$ which
satisfies the Killing equation $\nabla_{\mu} \xi_{\nu} + \nabla_{\nu} 
\xi_{\mu}=0$.  Using all existing symmetries, and complementing any missing 
vectors by appropriate coordinate basis vectors, we introduce four linearly
independent vectors $\xi^{\nu}_{(a)}$, with $a$ running from zero to
three. Upon defining the contracted vectors $K^{\mu}_{(a)}=
\xi^{\nu}_{(a)} T^{\mu}_{\nu}$, one obtains instead
of~(\ref{eq:stress-energy}),
\begin{equation}
\nabla_{\mu} K^{\mu}_{(a)} = T^{\mu}_{\nu} \nabla_{\mu}
\xi^{\nu}_{(a)} \, ,
\label{eq:killing-form}
\end{equation}
which becomes a true conservation law $\nabla_{\mu} K^{\mu}_{(a)} =
0$, for each $\xi^{\nu}_{(a)}$ that satisfies the Killing equation.
If $\xi^{\nu}_{(a)}$ is an {\em approximate} Killing vector in some
direction, then the corresponding current would be approximately
conserved $\nabla_{\mu} K^{\mu}_{(a)} \approx 0$ and the related
source term would capture the, possibly very small, deviations from
pure conservation.

The state vector ${\bf U} = \left[{\bf U}_{(a)},{\bf U}_4 \right]^{T}$
in this case is given by
\begin{eqnarray}
{\bf U}_{(a)} & = & \vol \xi^{\nu}_{(a)} T^{0}{}_{\nu} 
= \vol (\rho h u^{0} \xi^{\nu}_{(a)} u_{\nu} + p \xi^{0}_{(a)}) \, , \\
{\bf U}_{4}   & = & \vol J^{0}  = \vol \rho u^{0} \, ,
\label{eq:conserved_xi}
\end{eqnarray}
with the flux vectors given by
\begin{eqnarray}
{\bf F}^{j}{}_{(a)} & = & \vol \xi^{\nu}_{(a)} T^{j}{}_{\nu} 
= \vol (\rho h u^{j} \xi^{\nu}_{(a)} u_{\nu} + p \, \xi^{j}_{(a)} )\, , \\
{\bf F}^{j}{}_{4}   & = & \vol J^{j}    = \vol \rho u^{j} \, ,
\label{eq:fluxes_xi}
\end{eqnarray}
and the geometric source terms are
\begin{eqnarray}
{\bf S}_{(a)} & = &   T^{\mu}{}_{\nu} \nabla_{\mu} \xi^{\nu}_{(a)} \, , \\
{\bf S}_{4} & = & 0 \, .
\label{eq:sources_xi}
\end{eqnarray}

The linear relation between the standard form of the equations and one
adapted to Killing symmetries is given simply by the matrix
\begin{equation}
{\bf G} = \left[
\begin{array}{cc}
 \xi_{\nu(a)} &      0    \\
 0            &      1
\end{array} \right] \, .
\end{equation}

\subsubsection{Optimal source terms for RHD in general spacetimes}

The considerations in the last paragraph lead naturally to the case
also important for {\em numerical relativity}, namely the case of no
symmetries. Upon introducing a tetrad adapted to the coordinates, and
using $K^{\mu}_{(a)}= \epsilon^{\nu}_{(a)} T^{\mu}_{\nu}$, the
form~(\ref{eq:killing-form}) results to
\begin{equation}
(\sqrt{-g}\, T^{\mu}{}_{\nu})\,_{,\mu}  = - \Delta^{\rho}_{\mu\nu}
(\sqrt{-g} \, T^{\mu}{}_{\rho}) \, ,
\label{eq:mixed-form}
\end{equation}
where
\begin{equation}
\Delta^{\rho}_{\mu\nu}  = \frac{1}{2}  g_{\mu\lambda}
g^{\lambda\rho}{}_{,\nu} \, .
\end{equation}

With the definition of the state vector according to
\begin{eqnarray}
{\bf U}_{\mu} & = & \vol T^{0}{}_{\mu}  = 
\vol (\rho h u^{0} u_{\mu} + p \delta^{0}_{\mu}) \, , \\
{\bf U}_{4} & = & \vol J^{0}  = \vol \rho u^{0} \, ,
\label{eq:conserved_B}
\end{eqnarray}
the flux vectors
\begin{eqnarray}
{\bf F}^{j}{}_{\mu} & = & \vol T^{j}{}_{\mu} = 
\vol (\rho h u^{j} u_{\mu}  + p \delta^{j}{}_{\mu} ) \, ,\\
{\bf F}^{j}{}_{4} & = & \vol J^{j}  = \vol \rho   u^{j} \, ,
\label{eq:fluxes_B}
\end{eqnarray}
and the source terms
\begin{eqnarray}
{\bf S}_{\mu} & = & - \, (
\Delta^{0}_{0\mu} {\bf U}_{0} + \Delta^{k}_{0\mu} {\bf U}_{k}
+ \Delta^{0}_{k\mu} {\bf F}^{k}_{0} + \Delta^{k}_{l\mu} {\bf F}^{l}_{k} ) \,, \\
{\bf S}_{0} & = & 0 \, ,
\label{eq:sources_B}
\end{eqnarray}
we obtain a second form for the equations of RHD. Note that for notational
economy we have used a lowercase state vector index in the above
expressions.  With the explicit substitution of the perfect fluid stress-energy
tensor, we obtain
\begin{equation}
{\bf S}_{\mu}  =  \frac{\vol}{2} \rho h u^{\lambda}
u^{\rho} g_{\lambda\rho}{}_{,\mu}  - p (\vol)_{,\mu}  \, . 
\end{equation}

The source terms in this case involve considerably fewer summations of
metric derivatives compared to the standard
form~(\ref{eq:sources_A}). The mixed form
$\nabla_{\mu}T^{\mu}{}_{\nu}=0$ has been used before as the starting
point for non-conservative approaches to the RHD
equations~\cite{wilson72}. We point out here that the the related
source simplification should benefit conservative formulations as
well. The relation of the two forms is of course a lowering of the
free spacetime index, which is captured by a linear transformation of
the form
\begin{equation}
{\bf G} = \left[
\begin{array}{cc}
 g_{\mu\nu} &      0    \\
 0          &      1
\end{array} \right] \, .
\end{equation}

We also include here the source terms in the notation of the ``3+1"
spacetime decomposition~\cite{MTW}, which is a commonly adopted starting 
point for developing algorithms for dynamical evolutions of spacetimes. Assuming
a foliation with spacelike surfaces having a unit normal $n^{\mu}$,
the four-metric is decomposed as
\begin{equation}
g_{\mu\nu} = \gamma_{\mu\nu} - n_{\mu} n_{\nu} \, ,
\end{equation}
where $\gamma_{ij}$ is the 3-metric of the hypersurfaces.  Upon
introducing the lapse function $N$ and the spacelike shift
vector $N^i$~\cite{wald}, the evolution proceeds along the vector field $t^{\mu} =
N n^{\mu} + N^{\mu}$ and the components of the metric read explicitly
\begin{eqnarray}
g_{00} & = & - N^2 + \gamma_{ij} N^i N^j, \\
g_{0j} & = & \gamma_{ij} N^j, \\
g_{ij} & = & \gamma_{ij} \, .
\end{eqnarray}

The state vector we use differs from the usual
one (see, e.g.,~\cite{wilson72,font-washu}) in that it does not explicitly use the
above decomposition. For example, ${\bf U}^0 = T^{00}$ in contrast to
$\rho_{H} = T^{\mu\nu} n_{\mu} n_{\nu}$. The state vector variables
hence lose their usual meaning as observables in the instantaneous
Eulerian rest frame. In a generic spacetime this frame is of no
special significance and simply reflects the particular choice of
lapse and shift vector. In compensation, our choice of state vector is
valid in the absence of a spacelike foliation.

The volume element now reduces to $\sqrt{-g}=N\sqrt{\gamma}$, where
$\gamma$ is the determinant of the 3-metric, and the source terms for
equation~(\ref{eq:mixed-form}) have the following explicit form:
\begin{eqnarray}
{\bf S}_{\mu}  = &   
- \sqrt{\gamma} N^2 N_{,\nu} (u^{0})^2 \rho h  \nonumber \\ 
 & + \frac{1}{2} N \sqrt{\gamma} \rho h \gamma_{ij,\nu}
\left[(u^{0})^2 N^i N^j + u^i u^j + 2 u^0 u^i N^j \right] \nonumber \\ 
 & + N \sqrt{\gamma} u^0 \rho h \gamma_{ij} N^i{}_{,\nu} \left[u^0 N^j + u^j\right].
\end{eqnarray}

\section{Further analysis of the RHD equations}
\label{sec:analysis}

In contrast to non-relativistic hydrodynamics, the relativistic theory
exhibits a non-linear algebraic coupling of all equations (including
the continuity equation) through the velocity normalization
condition. This feature has wide ranging implications for the
structure of the theory, and in particular:
\begin{itemize}
\item The computation of fluxes from the state vector. In
non-relativistic hydrodynamics, the fluxes can be written as explicit
functions of the state vector. In RHD both fluxes and state vector are
more properly seen as algebraic functions of suitable ``primitive''
variables, as is obvious e.g., from 
equations~(\ref{eq:conserved_A1}-\ref{eq:fluxes_A}).
\item The analysis of the local characteristic structure.  The implicit
dependence of the fluxes on the state vector in RHD requires, in turn, that the
analysis of the Jacobians of the fluxes with respect to the state
vector be done through a set of intermediate variables.
\end{itemize}

A unique set of such intermediate variables employed in both procedures is
commonly used in the literature without explicit statement. There are few
constraints on the choice of primitive variables ${\bf w}$. An appropriate choice
influences the algebraic difficulty of analyzing the characteristic
structure of the system, a key ingredient of state-of-the-art Riemann solver
based numerical schemes for non-linear conservation laws~\cite{leveque}.
The rest mass density $\rho$ is a common candidate in all
proposed sets~\cite{Falle,betal97}. With our choice of representation
(e.g., equations~(\ref{eq:upper}) or~(\ref{eq:mixed-form})), the most
appropriate velocity variable is $u^{i}$, although a lower index would
lead to simpler calculations in the latter case. The
set must be completed with the choice of an additional thermodynamical
variable. Both the specific internal energy~\cite{betal97} and
pressure~\cite{Falle} have been used. The choice of pressure leads
to slightly simpler analysis. A reasonable compromise between
simplicity and maintaining continuity with past work is hence the
choice ${\bf w}=(\rho,u^i,\varepsilon)$.

\subsection{Characteristic structure}

Using the intermediate variables ${\bf w}$ the conservation
law~(\ref{eq:cons-law}) is rewritten as a quasi-linear system
\begin{equation}
{\bf A}^{0} {\bf w}{}_{,x^{0}}
+  {\bf A}^{j} {\bf w}{}_{,x^{j}} = 0 \, ,
\label{A-system}
\end{equation}
where
\begin{equation}
{\bf A}^{0} = \frac{\partial{\bf U}}{\partial {\bf w}}, \, \, \, \, \, \,
{\bf A}^{j} = \frac{\partial{\bf F}^{j} }{\partial {\bf w}} \, .
\end{equation}

Hence, upon introducing the intermediate eigenvalues $\hat{\lambda}^{j}$
\begin{equation}
det({\bf A}^{j} - \hat{\lambda}^{j} {\bf A}^{0}) =  0 \, ,
\end{equation}
and the intermediate right and left eigenvectors,
\begin{eqnarray}
& & ({\bf A}^{j} - \hat{\lambda}^{j} {\bf A}^{0}) \hat{{\bf r}}^{j} = 0\\
& & \hat{{\bf l}}{}_{j} ({\bf A}^{j} - \hat{\lambda}^{j} {\bf A}^{0})  = 0 \, ,
\end{eqnarray}
elementary algebra establishes that $\hat{\lambda}^{j} = \lambda^{j}$
and the right eigenvectors of the matrix ${\bf B}^{j}$ are given by
${\bf r}^{j} = {\bf A}^{0} \hat{\bf r}^{j}$.  The left eigenvectors
of this matrix are given simply by $\hat{{\bf l}}_{j} = {\bf l}_{j}$.

The  ${\bf A}^{\mu}$ matrices for the equations of RHD are
\begin{equation}
{\bf A}^{0} = \left[
\begin{array}{lll}
Y u^0 u^0 + \kappa g^{00} &  2 \rho h \mu_i  u^0                 &  Z u^0 u^0  + \chi g^{00} \\
Y u^k u^0 + \kappa g^{k0} & \rho h (\delta^k_i u^0 + u^k \mu_i)  &  Z u^k u^0  + \chi g^{k0}\\
    0                                   & \rho \mu_i                           & u^0
\end{array} \right] \, ,
\end{equation}
\begin{equation}
{\bf A}^{j} = \left[
\begin{array}{lll}
Y u^0 u^j + \kappa g^{0j} & \rho h (\mu_i u^j + u^0 \delta^j_i)      & Z u^0 u^j + \chi g^{0j}  \\
Y u^k u^j + \kappa g^{kj} & \rho h (\delta^k_i u^j + u^k \delta^j_i) & Z u^k u^j + \chi g^{kj}  \\
 0                                      & \rho \delta^j_i                          & u^j
\end{array} \right] \, ,
\end{equation}
where $Y=\rho + \kappa$, $Z=1 + \varepsilon + \chi$ and
\begin{equation}
\mu_i  =  \frac{\partial u^0}{\partial u^i}=-\frac{u_i}{u_0} \, , \, \, \,
\chi  =  \frac{\partial p}{\partial\rho} \, , \, \, \,
\kappa  =  \frac{\partial p}{\partial\varepsilon} \, .
\end{equation}

We choose a coordinate direction, which we label `1'.  The other two
coordinate directions are then denoted by $A=(2,3)$.

The matrix ${\bf A}^1- \lambda^{1} {\bf A}^0$ has {\it eigenvalues}
\begin{equation}
\lambda^{1}_0 = v^1 \mbox{\,\,\,\,(triple)} \, ,
\label{lambda0}
\end{equation}
and
\begin{equation}
\lambda^{1}_{\pm} = \frac{1}{1-c_{s}^{2}(1+L)}
\left[ - M c_{s}^2 + v^1 (1-c_{s}^{2}) \pm
c_s \sqrt{D} \,\right] \, ,
\label{lambdapm}
\end{equation}
where
\begin{equation}
D =  c_s^2 (M^2 - LN) + (1-c_s^2) (N - 2 M v^1 + L  (v^1)^2) \, , 
\end{equation}
with $v^1 = \frac{u^1}{u^0}$. The local sound
speed is denoted by $c_s$ and satisfies
\begin{equation}
h c^2_s = \chi +\frac{p}{\rho^2}\kappa \, ,
\end{equation}
and the following shorthand notation was also used:
\begin{equation}
L=\frac{g^{00}}{(u^0)^2} \, ,
M=\frac{g^{01}}{(u^0)^2} \, ,
N=\frac{g^{11}}{(u^0)^2} \, .
\end{equation}

A complete set of linearly independent {\it right-eigenvectors} 
(${\bf r}^1 = {\bf A}^0 \hat{\bf r}^1$), is given by
\begin{equation}
{\bf r}_{0,1} = \left[ u^{\mu},\frac{1}{\alpha} \right]^{T} \, ,
\end{equation}
\begin{equation}
{\bf r}_{0,A} = \left[ \delta^{\mu}_A (1 + u^{B}u_{B}), 0 \right]^{T}
+ u_{A} \left[u^0, u^1,0,0,\frac{1}{h}-\frac{1}{\alpha}\right]^{T} \, ,
\end{equation}
\begin{equation}
{\bf r}_{\pm} = \left[ u^{\mu} + \frac{\Lambda_{\pm}}{(u^{0})^{2}}
(u^{1}g^{0\mu}-u^{0}g^{1\mu}) , \frac{1}{h} \right]^{T} \, ,
\end{equation}
with the definitions
\begin{eqnarray}
\alpha  &\equiv& 1+\varepsilon-\frac{\chi}{\kappa}\rho \, , \\
\Lambda_{\pm}    &\equiv& \frac{c^{2}_{s}}{
(v^{1}-\lambda_{\pm}) (1-c^{2}_{s}) - c^{2}_{s} (M - \lambda_{\pm} L)} \, .
\end{eqnarray}

We note that the ${\bf r}_{\pm}$ eigenvectors are unique up to
normalization, whereas the ${\bf r}_{0,1},{\bf r}_{0,A}$ vectors can
be any set spanning the degenerate subspace. For a perfect fluid EOS
these expressions coincide with the ones reported in~\cite{pf99}. Note
that in that restricted case $\alpha=1$. The spectral decomposition
given above applies to a chosen direction $j$. Since $j$ is arbitrary,
to obtain similar expressions for the remaining directions, it
suffices to specialize them accordingly, e.g., obtain the eigenvalues
from expressions~(\ref{lambda0}) and (\ref{lambdapm}) with
substitution of the desired direction, and permutation of the
corresponding eigenvectors. Complete sets of eigenvectors for other
versions of the equations are obtained by a straightforward
multiplication with the corresponding ${\bf G}$ matrix and are not
reproduced here.

\subsection{Inverting the state vector}

The general statement of the recovery of the primitives fields is to
find values for e.g., ${\bf w}=(\rho,\varepsilon,u^{i})$, given a set of
conserved variables ${\bf U}$. This procedure is part of any solution algorithm
that uses the conservation form of RHD, i.e., it is not related to the
characteristic decomposition of the system.

We focus on the standard form of the equations, in which case ${\bf
U}=\vol (T^{00},T^{0i},J^{0})$. At first sight, this requires the
inversion of the system of six non-linear algebraic equations
\numparts
\begin{eqnarray}
\label{eq:nra}
T^{00} & = & \frac{1}{\rho} (\frac{p}{\rho} + 1 + \varepsilon) 
(J^0)^2 + p g^{00} \, , \\ 
T^{0i} & = & ( \frac{p}{\rho} + 1 + \varepsilon) J^0 u^i + p g^{0i} \, , \\
 - \rho^2 & = & g_{00} (J^0)^2 + 2 \rho g_{0i} J^{0} u^i 
+ \rho^2 g_{ij} u^i u^j, \\  
p   & = & p(\rho,\varepsilon) \, .
\label{eq:nrd}
\end{eqnarray}
\endnumparts

We point out that even in the most general case the size of the
non-linear system can be reduced, with the elimination of the velocity
from the unknowns. We introduce the tensor
$S^{\mu\sigma}=g_{\nu\rho}T^{\mu\nu}T^{\rho\sigma}$.  Inspection of
the $S^{00}$ component of this tensor shows immediately that it is
only a function of conserved variables.  Together with $T^{00}$ and
the EOS, we have a reduced system which reads
\numparts
\begin{eqnarray}
\label{eq:nr2a}
S^{00} & = & (\frac{p}{\rho}+1+\varepsilon)(\frac{p}{\rho}-1-\varepsilon)
(J^{0})^2 + p^2 g^{00} \, ,  \\ 
T^{00} & = &  \frac{1}{\rho}
(\frac{p}{\rho}+1+\varepsilon) (J^{0})^2 + p g^{00} \, , \\ 
p & = & p(\rho,\varepsilon) \, .
\label{eq:nr2c}
\end{eqnarray}
\endnumparts

The use of a general EOS which may be only available in tabulated form
implies that the reduced system~(\ref{eq:nr2a}-\ref{eq:nr2c}) is in general to
be inverted numerically with an iteration procedure. A description of
a typical iterative procedure (applied to equations~(\ref{eq:nra}-\ref{eq:nrd}
in the special relativistic limit) is
given in~\cite{genesis}, where estimated values for
$(\varepsilon,v^i,\rho)$ are used to start a non-linear iteration of
the system. In~\cite{em95} several other procedures are discussed for
the special case of a polytropic gas.

It is of interest to point out that further reduction of the
system to a binomial equation, and hence with an explicit and convenient
solution, is possible for special classes of analytic EOS.  Assuming
that the EOS if of the explicit form $p=\rho F(h)$ allows the further
manipulation of~(\ref{eq:nr2a}-\ref{eq:nr2c}) to obtain
\begin{equation}
\fl
(T^{00})^2 S^{00} - h (2 F(h) - h) (J^0)^2 (T^{00})^2 
- (S^{00} - h (F(h) - h) (J^0)^2)^2 g^{00} = 0 \, , 
\label{eq:nr-final}
\end{equation}
which is a non-linear equation for the enthalpy.  An explicit relation
between conserved and primitive variables, which rests on the ability
of solving equation~(\ref{eq:nr-final}), has an impact on the
efficiency of the numerical code, as it eliminates an iterative
process that is required, at least once per each spacetime point.

It is seen immediately that the metric component $g^{00}$ stands out
as having special significance in this algebraic equation. Indeed, as
already pointed out in~\cite{pf99} any null foliation (characterized
by $g^{00}=0$) leads to explicit solutions for $h$ in terms of
conserved quantities, in the case of a polytropic equation of state $p
=(\Gamma-1) \rho \varepsilon$, where $\Gamma$ is the adiabatic index. 
The most general case explicitly
reducible to a binomial is $F(h)= \alpha/h + \beta + \gamma h$, where
$(\alpha,\beta,\gamma)$ are constants characterising the fluid. The
polytropic gas is the special case with $\alpha=0,\gamma = -\beta =
(\Gamma-1)/\Gamma$.

For spacelike foliations ($g^{00}\not=0$), the situation is slightly
different. The choice $F(h)= \alpha/h + \beta  + h$ leads to a
binomial for $h$, whereas the choice $F(h)= \alpha /h + \gamma h$ leads
to a binomial for $h^2$. None of those cases includes the 
polytropic gas, but see the arguments of~\cite{zss} for an example of
the latter case, with $\gamma = - \alpha = 1/4$.

Once the enthalpy is obtained, the other variables follow
straightforwardly, e.g., the velocity follows from
\begin{equation}
u^i = \frac{\rho(T^{0i} - p g^{0i})}{(p + \rho + \rho \varepsilon)
J^0} \, .
\end{equation}

\section{Summary and concluding remarks}
\label{sec:summary}

In view of the increasing importance of conservative RHD in numerical
applications in relativistic astrophysics and relativity, we explored
the corresponding framework in some detail.  The algebraic complexity
of the systems involved suggests that links between specific
manifestations of the equations be established at the outset. This is
accomplished with the introduction of simple linear transformations of
the equations, which leave the local characteristic structure
invariant but have non-trivial impact on the source terms.  It is
pointed out that for spacetimes with exact or approximate symmetries,
but also in the general case, the equations can be modified to capture
the conservation property in an optimal way.

The local characteristic structure of the RHD equations in a general
spacetime foliation (spacelike or lightlike) is analyzed for a general
equation of state (EOS). This extends previous results restricted to
the polytropic case.  Special classes of EOS are identified for both
spacelike and null foliations, which lead to explicit inversion of the
state vector and computational savings. In a lightlike foliation, the
commonly used polytropic EOS is included in the explicitly invertible
cases.

Conservative RHD techniques have been applied only recently to general
3D spacetime evolutions~\cite{font-washu}. It seems worthwhile to
investigate the benefits of the increased control over source term
structure presented here, in situations of current interest, e.g., the
study of neutron star binary coalescence.

The presented framework is uniquely suitable for strong-field
simulations using {\it lightlike} foliations of the
spacetime. Lightlike foliations attached to the exterior of spacelike
surfaces have been suggested as an effective way for providing global
spacetime solutions~\cite{nigel-ccm} (for a review
see~\cite{LR-jeff}). The whole approach has been dubbed
Cauchy-characteristic matching (CCM). Fluid evolution in the CCM
context has already been investigated~\cite{ddv}. The RHD framework
proposed here is the natural candidate for providing economical and
state of the art numerical fluid evolution in both foliations, as it
is form invariant with respect to the slicing.  Algorithmically, the
only change between spacelike/null domains would be the routine used
for the recovery of primitives.

A recent demonstration of long-term stable numerical evolutions of a
single black hole based on null coordinates~\cite{60000M} opens the
possibility for applications to non-vacuum single black hole
environments solely within a lightlike framework. A full
implementation of the present formulation in the case of spherical
symmetry has been presented~\cite{pf99}, illustrating the ease of
studying black hole growth through accretion using an ingoing null
foliation. Three dimensional implementations for fixed black hole
backgrounds in null coordinates are also available and results will be
discussed elsewhere.

\ack P.P. acknowledges support from the AEI, Golm while part of this
work was completed. We thank J.M$^{\underline{\mbox{a}}}$. Ib\'a\~nez for
many stimulating discussions during the course of this work.

\end{document}